# A clean, bright, and versatile source of neutron decay products

D. Dubbers[1], H. Abele[1], S. Baeßler[2], B. Märkisch[1], M. Schumann[1], T. Soldner[3], O. Zimmer[3,4]

[1]Physikalisches Institut der Universität, Philosophenweg 12, D-69120 Heidelberg, Germany
[2]Institut für Physik, Johannes Gutenberg-Universität, 55099 Mainz, Germany
[3]Institut Max von Laue-Paul Langevin, BP 156, F-38042 Grenoble Cedex 9, France
[4]Technische Universität München, D-85748 Garching, Germany

(Dated: September 27. 2007)

*Abstract*: We present a case study on a new type of cold neutron beam station for the investigation of angular correlations in the β-decay of free neutrons. With this beam station, called PERC, the 'active decay volume' lies inside the neutron guide, and the charged neutron decay products are magnetically guided towards the end of the neutron guide. Hence, the guide delivers at its exit a beam of decay electrons and protons, under well-defined and precisely variable conditions, which can be well separated from the cold neutron beam. In this way a general-purpose source of neutron decay products is obtained which can be used for various different experiments in neutron decay correlation spectroscopy. A gain in phase space density of several orders of magnitude can be achieved with PERC, as compared to existing neutron decay spectrometers. Neutron beam related background is separately measurable in PERC, and magnetic mirror effects on the charged neutron decay products and edge effects in the active neutron beam volume are both strongly suppressed. Therefore the spectra and angular distributions of the emerging decay particles will be distortion-free on the level of $10^{-4}$, more than 10 times better than achieved today.

*PACS:* 13.30.Ce, 23.40.Bw, 28.52.Lf, 29.27.Fh

*Keywords:* Cold neutrons, neutron guide, beta decay, weak interaction

## 1. Introduction

In free neutron decay, a large number of measurable quantities are accessible or in reach today. These are, besides the neutron lifetime $\tau$, the electron-antineutrino correlation coefficient $a$, the Fierz interference term $b$, the parity P-violating β-asymmetry $A$, antineutrino asymmetry $B$, and proton asymmetry $C$, the time-reversal T-violating triple correlation coefficient $D$, and various correlation coefficients involving the longitudinal polarization of β-particles, as well as the β- and proton energy spectra associated with these quantities.

The Standard Model of particle physics, on the other hand, requires essentially only two parameters to describe neutron decay (shortly n-decay), namely the quark mixing CKM matrix element $V_{ud}$ and the ratio of axial vector to vector coupling constants $\lambda = g_A/g_V$, if the Fermi coupling constant $G_F$ is taken from muon decay, and if T-conservation is assumed. Hence, the problem is strongly over-determined, and there is ample room for precision checks for physics beyond the Standard Model, for instance on CKM non-unitarity, right-handed amplitudes, scalar or tensor amplitudes, and others, as required in many extensions of the Standard Model, see for instance the reviews [1-6]. Neutron-decay, via induced terms, may also give information on QCD, and, furthermore, plays an important role in astrophysics and cosmology. Hence, in n-decay studies, important questions of particle and astrophysics can be addressed, and, like in the high-energy sector, these investigations should be done as precisely as possible. For these reasons free n-decay is a very active field, and many new projects are underway worldwide [7].

Conventionally, 'in-beam' n-decay spectroscopy uses a section ('active volume') of a free-flying n-beam after this beam has left the neutron guide (or beam tube), and the n-decay products, electrons and/or protons (shortly e/p), are registered in some suitable e/p-detectors positioned next to the n-beam. Existing spectrometers for the measurement of n-decay correlation coefficients are PERKEO (in its versions II and III, as specified below), built for the measurement of *A*, *B*, and *C* [8-11], and *a*SPECT, built for the measurement of *a* [12]. In addition, there is a number of new n-decay spectrometers projected or under construction, for an overview see [7].

In this paper we investigate the possibility to place this active decay region into the interior of a neutron guide (shortly n-guide), and to have the decay particles emerge through the open end of the guide and be guided to an e/p-detector installed outside the n-guide. In the proposed beam station, called PERC (short for Proton and Electron Radiation Channel) the charged decay products are collected by a strong magnetic field $B_0$ applied along the n-beam axis *z*, and are magnetically guided towards the end of the n-guide, where they can be separated from the n-beam and can be further investigated.

The use of a strong magnetic e/p-guiding field along the n-beam has some great advantages. The magnetic field will project the beam of n-decay products emerging from the large active volume onto the central region of some e/p-detector of relatively small size. Furthermore, n-spin always adiabatically follows the local field direction ***B***(***x***), and there will be a clear-cut division between decay particles (momentum ***p***) emitted along the local field direction (***p·B*** > 0) and decay particles emitted against the local field direction (***p·B*** < 0), so the solid angle of e/p-detection no longer depends on the geometry under which the e/p-detector sees the n-decay volume.

In short, with PERC, the n-guide will deliver at its exit not a flux of neutrons, but a flux of charged n-decay products under well-defined conditions, ready to be used for a larger number of different n-decay angular correlation experiments. On first sight, such a device appears to have many inherent sources of experimental error, but we shall show in Section 2 how these errors can be circumvented by rather simple means. In Section 3 we list some typical applications of PERC, and in Section 4 we thoroughly discuss the level of remaining experimental uncertainties, which are summarized in Table 1. − As PERC will require a larger investment, this paper is intended, beforehand, to instigate public discussion of this project.

## 2. Principle of PERC

The number of neutrons that decay *inside* a cold n-guide (of typical cross-section 100 cm$^2$) is surprisingly large, of order $10^6$ per second and per meter length of guide for a continuous reactor neutron source like ILL in Grenoble or FRM-II in Munich (see Appendix 1). If the 'in-guide' decay volume is chosen sufficiently long, and the area of the e/p-detectors is kept sufficiently small, then a large gain in phase space density of detected decay products will be possible, which could strongly enhance the quality of the mostly statistics limited ongoing experiments, and will also bring previously inaccessible n-decay observables into reach.

Before going into further details let us first list the phase space operations that we wish to perform on such a beam of charged decay particles with momentum *p*, angle of emission *θ* relative to the n-beam axis (= axis of magnetic field $B_0$), and $\sin \theta = p_x/p$:

*Operation I*: limit the *divergence* of the emerging e/p-beam by some freely variable cut-off angle $\theta < \theta_c$ without changing any other e/p-beam properties;
*Operation II*: limit the *width x* of the e/p-beam without changing any other e/p-beam properties;
*Operation III*: trade e/p-beam *width x* against e/p-beam *divergence* $\theta_c \sim p_x$ under well defined conditions.

It will turn out in the following that a beam of n-decay products with these properties will be extremely versatile and will permit to measure a larger number of basic n-decay parameters with high statistics. In addition, the resulting e/p-beam has several interesting features which will reduce also systematic errors far below what is possible with existing n-decay spectrometers.

*2.1 Limitation of e/p-beam divergence without spectral distortions*

*Operation I*, the introduction of a cut-off angle $\theta_c$ to e/p-transport, will be performed by having the e/p-beam pass a region of strongly enhanced field $B_1 \gg B_0$ before it reaches the detector, see Fig. 1. This field will act as a magnetic mirror and will transmit only a fraction of $\approx B_0/4B_1$ of all decay particles, namely those emitted downstream under angles below the cut-off angle $\theta_c$ (Appendix 2). Their initial angle of emission $\theta_0$ in field $B_0$ hence is limited to

$$\theta_0 \leq \theta_c = \arcsin\sqrt{B_0/B_1}. \tag{1}$$

The cut-off angle $\theta_c$ can be tuned by varying the ratio $B_0/B_1$ (cf. Fig. 3a).

Further downstream of $B_1$, the magnetic field $B(z)$ decreases again. When $B(z)$ again reaches the starting value $B_0$, the e/p-beam will have regained its original width $x_0$, but will be limited to angles $\theta < \theta_c$, which is Operation I. All field variations $B(z)$ must be slow enough such that the electrons and protons are transported adiabatically. In this case magnetic transport no longer depends on the energy $E$ of the charged particles.

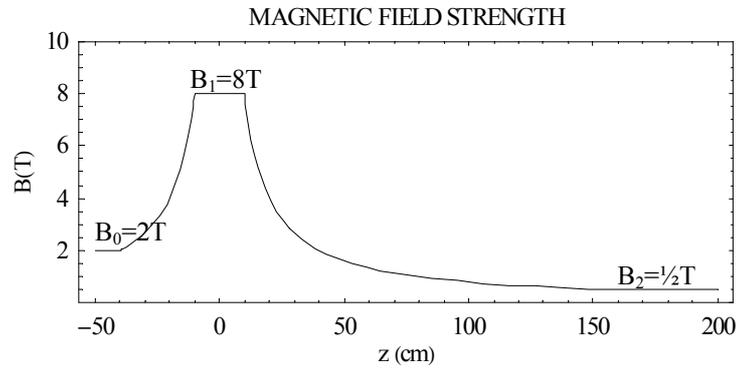

Fig. 1: Sketch of magnetic field $B(z)$ along the central field line, in 'standard field configuration'. Neutron decay electrons and protons enter from the left, from inside the neutron guide (cf. Fig. 2), and are detected on the far right.

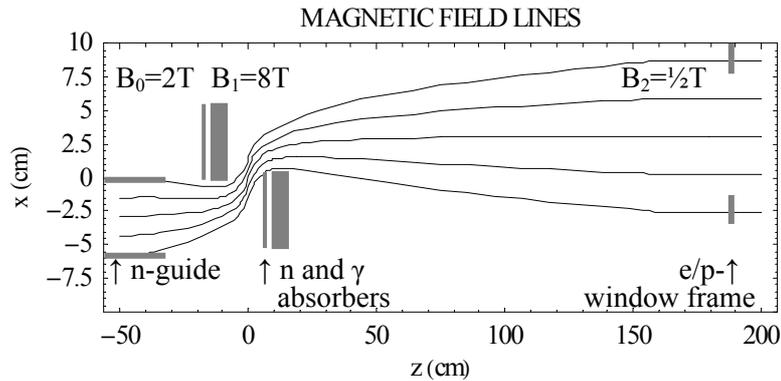

Fig. 2: A possible scheme for the separation of the neutron beam from the electron and proton beam. Shown are the $B$-field lines near the exit of the neutron guide. Note that the vertical scale is 5 times the horizontal scale, the maximum slope of the field lines is only $10^0$. Neutron decay products enter from inside the neutron guide and are magnetically guided to the detector through the e/p-window frame, as described in the text. Other methods of beam separation are also possible.

*Operation III*, the trade of width against divergence of the e/p-beam, can be done by variation of the final field $B_2$, because for adiabatic transport of charged particles in a magnetic field $B$, the quantity $a^2B$ is a constant of the motion, where $a$ is the radius of gyration. Radius of gyration $a_2$ at field value $B_2$, and with it the total width $2x_2$ of the e/p-beam, and divergence $\theta_2$ therefore scale as

$$\frac{a_2}{a_0} = \frac{x_2}{x_0} = \frac{\sin\theta_0}{\sin\theta_2} = \sqrt{\frac{B_0}{B_2}}. \tag{2}$$

Hence, e/p-beam width $2x_2$ can be traded for e/p opening angle $2\theta_2$, and vice versa, by varying $B_2$. When the e/p are emitted in field $B_0$ with momentum $p$, the maximum radius of gyration for $\theta_0 = 90^0$ is $r_0 = p/(eB_0)$. For the maximum momentum occurring in n-decay, $p_{max} = 1.20$ MeV/c, this radius is limited by $r_0 \leq r_{0\,max}(\text{mm}) = 4.0/B_0(\text{T})$, both for the decay protons and electrons. The conditions for adiabatic transport will be discussed in Section 4.2.

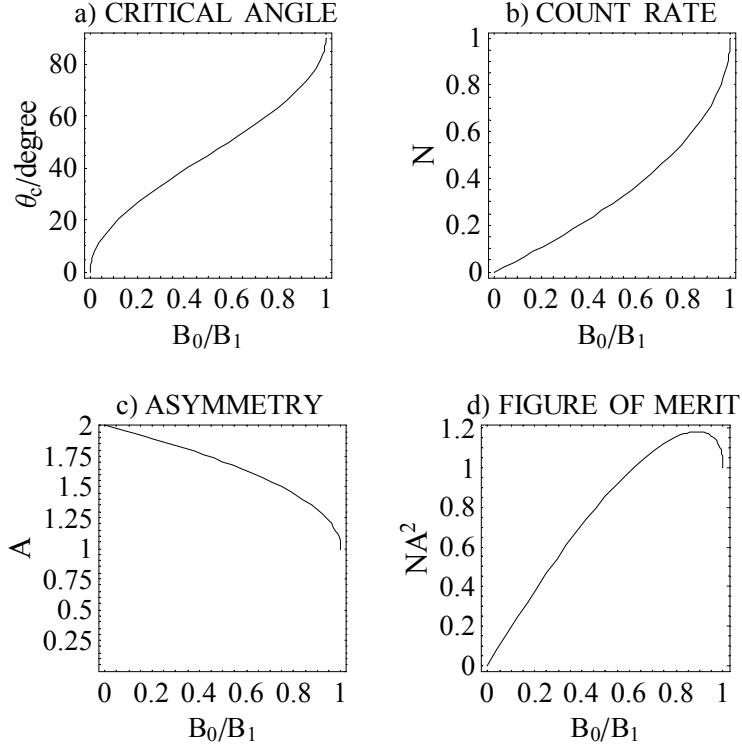

Fig. 3: Magnetic mirror effect on the e/p beam from n-decay, in dependence of $B_0/B_1$. $B_0$ is the field in n-decay region, $B_1$ is the mirror field, cf. Fig. 1.
Shown are the field dependences of:
    a) critical angle $\theta_c$ of magnetic mirror reflection, Eq. (1);
    b) count rate $N$ of decay particles, Eq. (A10);
    c) asymmetry signal $A_{exp}$ of decay particles, Eq. (A11);
    d) figure of merit $NA_{exp}^2$, Eq. (A12).
The last three curves are normalized to 1 for the case of no mirror field, $B_0/B_1 = 1$.

This variable magnetic field configuration $B(z)$ has other interesting features (Appendix 2) which will turn out to be extremely helpful:

Consider an e/p-angular distribution of type
$$W(\theta) = 1 + A^* \cos\theta, \tag{3}$$
with some experimental angular correlation signal $A^*$. In the case of the β-decay asymmetry $A$, this signal is $A^* = (v_e/c)\, AP_n$, with electron velocity $v_e$ divided by the speed of light $c$, and n-polarization $P_n$. Averaged over $\theta$ up to $\theta_c$, the experimental asymmetry becomes
$$A_{exp} = \tfrac{1}{2} A^* (1 + \cos\theta_c) \tag{4}$$
When there is no magnetic barrier, i.e. $B_1 = B_0$, the sensitivity of $A_{exp}$ to magnetic field inhomogeneities diverges, see Fig. 3c at $B_0/B_1 = 1$, and must be carefully dealt with, as was the case in previous instruments. When, as in PERC, a magnetic barrier $B_1 > B_0$ is applied, this sensitivity to field variations no longer diverges, and magnetic field inhomogeneities are much easier to handle. Further, with increasing $B_1$, one trades count rate $N$, Fig. 3b, against signal size $A_{exp}$, Fig. 3c, i.e. statistical against systematic precision. The figure of merit $NA_{\text{exp}}^2$, Fig. 3d and Appendix 2, permits a relatively wide range of values $B_0/B_1$. − The use of a magnetic barrier in n-decay was discussed independently in [13].

The values chosen for the magnetic field strengths $B_0$, $B_1$, $B_2$ will depend on the experiment to be performed. To be specific, we shall discuss the use of this beam station for field values chosen to $B_0 = 2$ T, $B_1 = 8$ T, and $B_2 = 0.5$ T, which we call the standard field configuration, as shown schematically in Fig. 1. The initial angles $\theta_0$ and radii of gyration $a_0$ at the starting point (in field $B_0$) of particles that reach the detector downstream then are limited to $\theta_0 \leq \theta_c = 30°$ and $a_0 = r_0 \sin\theta_0 \leq r_{0\,\text{max}} (B_0/B_1)^{1/2} = 1$ mm. The limits for the gyration radius $a$ and the pitch angle $\theta$ in these fields then are (Appendix 2):

| | | | |
|---|---|---|---|
| In the main n-decay field | $B_0 = 2$ T: | $a_0 \leq 1$ mm, | $\theta_0 \leq 30°$; |
| in the magnetic mirror field | $B_1 = 8$ T: | $a_1 \leq \tfrac{1}{2}$ mm, | $\theta_1 \leq 90°$; |
| in the detector field | $B_2 = \tfrac{1}{2}$ T: | $a_2 \leq 2$ mm, | $\theta_2 \leq 15°$. |

The cross section $bh$ of the n-guide is chosen to be square ($b = h$) with a width of $b = 6$ cm, the length of the active n-beam volume within the n-guide is chosen to $L = 7$ m (a length which still can be transported with reasonable means). In the following we shall call $b^2 \cdot L = 25$ liter the standard n-beam volume.

*2.2 Limitation of e/p-beam width without edge effects*

*Operation II*, the lateral confinement of the e/p-beam, needs closer investigation. If a decay particle e/p is emitted too close to one of the walls of the n-guide then the e/p will be absorbed on the guide wall, which leads to an unacceptable distortion of the e/p-spectra and angular distributions. With PERC this 'edge effect', due to lateral confinement of the decay volume, is more prominent than it is with PERKEO II [8], because the e/p leave the active n-decay volume through its smallest face, defined by the cross section of the n-guide, and not through its largest face like in PERKEO II, which has the e/p-collecting *B*-field at right angles to the n-beam. This effect can be eliminated if one installs, after the end of the guide, a *thin* e/p-absorbing frame which defines an open window for the decay particles (called e/p-window, shown on the right of Fig. 2), instead of a *thick* absorber with thickness of several gyration lengths like in the PERKEO spectrometers.

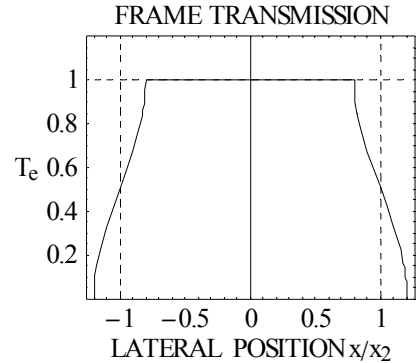

Fig. 4: Transmission function of the thin e/p-window frame, see also Figs. 2 and 5. The open window extends from $x/x_2 = −1$ to $x/x_2 = +1$. With this transmission profile, edge effects on the transmitted e/p-beam disappear for uniform n-density.

Let this e/p-window, positioned at field value $B_2$, have width $2x_2$ and height $2y_2$. If this window can be made sufficiently thin, then its transmission function $T_e$, for a given transversal e/p-momentum, is simply the fraction of the circumference of the gyration circle (radius $a_2$) which overlaps with the frame. Hence, in an $\pm a_2$ vicinity to the frame, $T_e$ depends on the $x$-position of the particle's guiding center as

$$T_e(x) = \frac{1}{\pi} \arccos \frac{|x| - x_2}{a_2}, \qquad (5)$$

as shown in Fig. 4, and similarly for the $y$-coordinate.

The disappearance of edge effects on this e/p-window is due to the following: Electrons or protons spiraling on a helical trajectory about $B_2$, close to the outer edge of the e/p-window, may or may not hit the window frame. But, interestingly, this effect does not distort the spectra or angular distributions of the decay particles: spiraling particles that hit the frame although their guiding center does not hit it, are, for a uniform n-beam profile, exactly compensated in number by those spiraling particles that sneak through the open e/p-window although their guiding center does hit the frame. This symmetry property of the transmission function $T_e$ relies on the fact that the small areas to the left and to the right of the vertical dashed lines in Fig. 4 are of the same size. Interestingly, the same effect can be obtained without any e/p-window when a position sensitive e/p-detector is used; in this case the lateral size of the decay volume can be chosen after the end of the measurement.

In the four corners of the window, this compensation of e/p-hits and misses is not complete, but this corner effect is well below $10^{-4}$. Further deviations from this symmetry, for instance due to a non-uniform n-flux in the decay volume, or due to the finite thickness of the e/p-window, will be discussed in Section 4.1, and backscattering off the e/p-frame in Section 4.4.

To be specific, we assume in the following that this e/p-window frame is just thick enough (~ 0.5 g/cm$^2$) to absorb all n-decay electrons of endpoint energy $E_{\beta\,max}$ = 783 keV. For an initial width $2x_0 = 2y_0 = 5$ cm of the e/p-beam, the size of this window will be $2x_2 = 2y_2 = 2x_0\,(B_0/B_2)^{1/2} = 10$ cm for standard field configuration, which we shall call the standard size of the e/p-window.

*2.3 Separation of the e/p-beam from the n-beam*

The e/p-beam which emerges through the open end of the n-guide can be guided magnetically (or, for the protons, also electrically) away from the guide and from the n-beam to any suitable place for further analysis. How this will be realized depends on the specific experiment to be performed with PERC. For the case that the emerging e/p-beam must remain collinear with the n-beam axis, we present, in Figs. 2 and 5, an example of how the e/p-beam can effectively be separated from the n-beam.

In our example the separation is done in the high field region $B_1$, where the diameter of the e/p-beam is compressed to a width $2x_1 = 2x_0\,(B_0/B_1)^{1/2} = 2.5$ cm for standard field configuration, i.e. the e/p-beam is confined to a rather narrow channel. We then position the n-beam stop near the maximum of the mirror $B_1$-field and let the e/p-beam escape through a suitable hole in the n-beam stop. The e/p-detector then can see the interior n-beam region at most through this small 'key hole' in the n-beam stop. If the $B_1$-field axis, and with it the hole in the n-beam stop, is made non-collinear with the n-beam axis, off by an angle of about $10^0$, then this channel is blocked for the neutrons. Equally blocked is the direct sight from the e/p-detector through this channel onto the n-decay region, while the charged e/p follow the curved magnetic field lines, as indicated in Fig. 2. $\mathbf{R}\times\mathbf{B}$-drifts of the charged e/p in the curved parts of the field $\mathbf{B}$ (with radius of curvature $\mathbf{R}$) can be neglected because $\mathbf{R}$ changes sign at $z = 0$ of Fig. 2. Beam related background for this configuration will be discussed in Section 4.3.

*2.4 Modes of operation*

In PERC, particle detection from n-decay is most easily done in *single-count* mode. For e/p-coincidence measurements (which are possible also with only one detector [9,10]) dead time in PERC would be too large in view of the high count rate and great length of the decay volume. But it turns out that, indeed, most observables in n-decay can be measured in single count mode, either with electrons or with protons, as will be discussed in Section 3.

PERC is most easily used with only one e/p-detector, installed behind the e/p-window, simultaneous e and p detection in two detectors, like in Fig. 6, being an exception. *a*SPECT [12] also uses only one detector, while PERKEO [8] is equipped with two detectors which cover both half-spaces, i.e. $|\theta_0| < 90^0$ and $|\theta_0| > 90^0$ with respect to the local $B_0$ field direction. In single count mode, a second detector may be helpful to eliminate fluctuations of n-intensity in polarized neutron work, but a highly stable in-beam n-monitor, installed upstream of the spectrometer, can do the same job. Further, in PERKEO, a second detector helps to diminuish magnetic mirror effects of the charged decay particles in a non-uniform $B_0(z)$ [14], but sensitivity to magnetic mirror effects will be much lower in PERC with one detector than it is in PERKEO with two detectors, see Section 4.2. Finally, the second detector is useful to reconstruct backscattered electron events, but this turns out not to be indispensable with PERC, see Section 4.4.

To compare PERC directly with previous instruments under equal conditions, we first consider the case when there is no mirror field, i.e. $B_1 = B_0 \geq B_2$. The e/p-count rate in a single detector downstream reaches $7 \cdot 10^5$ s$^{-1}$ for PERC operated with a continuous unpolarized n-beam and standard decay volume (Appendix 1). This rate is 1000 times higher than that of typical instruments with 'external' n-decay volumes like $a$SPECT [12] or PERKEO II [8-10], and more that 10 times that of our newly installed PERKEO III [15], which uses a free-flying n-beam like PERKEO II, but with a magnetic field directed along the external n-beam. PERKEO III is our high precision machine for the years to come until PERC goes into operation. Of course, for spectroscopic purposes this count rate in PERC of $7 \cdot 10^5$ s$^{-1}$ is far too high, and one can trade a large part of this rate for other purposes: For instance when, in the standard field configuration, we limit the primary angle of emission to $\theta_0 < \theta_c = 30^0$, then this count rate drops by a factor of nearly eight to $8 \cdot 10^4$ s$^{-1}$, Eq. (A3).

In addition, for $\theta_c = 30^0$, the correlation signals will be larger by a factor $1 + \cos \theta_c = 1.9$, as compared to the case of no mirror field $B_0 = B_1$, see Fig. 3c, so the figure of merit $NA^2$ of PERC will be about 400 times higher than that of PERKEO II, and 10 times that of PERKEO III. This is accompanied by another factor of 100 overall gain in phase space density, because in standard configuration, both e/p-beam size and e/p-emission angles are each ten times smaller in PERC than they are in the PERKEO instruments. This latter gain does not translate into improved statistics but into improved systematics: a smaller e/p-beam size leads to a smaller e/p-detector and with it to higher energy resolution and lower background, smaller e/p-emission angles lead to lower edge effects and lower backscattering, etc.

With PERC the n-beam can be used in *continuous* or in *pulsed* mode. In pulsed mode n-decay is observed from a moving n-cloud containing typically $I_n' \cdot \Delta t = 5 \cdot 10^8$ neutrons (Appendix 1). The count rate in pulsed mode is typically 12 times lower than the count rate in continuous mode, Eq. (A7). The n-pulse of about 2 m length moves with mean cold-neutron velocity $v_n = 800$ m/s through the $6 \times 6$ cm$^2$ square guide of length $L = 7$ m. The decay particles will be counted in the detector only while the n-pulse is fully contained within the magnetically uniform region of length $L' \approx 6$ m. With neutron time-of-flight data sampling, the size of the active volume can be selected even after the end of the experiment. We have tested the principle of n-decay from a propagating neutron cloud earlier with a free n-beam [16]. However, our discussion of error sources, Section 4, will show that even high-precision absolute measurements of n-decay correlation coefficients can be done in the continuous n-beam mode of PERC.

PERC can be operated with *polarized* or *unpolarized* cold neutrons. For absolute correlation measurements involving neutron spin, the highest achievable neutron polarization of, at present, $(99.7 \pm 0.1)$ % at 10 % polarizer transmission [17] (i.e. 10 % of the full *unpolarized* incoming n-flux) will be used. For measurements which involve neutron spin but which do not depend critically on the absolute degree of n-polarization, like weak magnetism, second class form factors, or radiative corrections (Section 3), a lower polarization of 98 % at 20 % transmission can be used. Like in the present 'fundamental physics' n-beam station at PF1B at ILL, all specialized neutron equipment including the PERC e/p-channel itself will be made easily removable. Examples of both polarized and unpolarized operation of PERC will be given in Section 3.

A setup of PERC for pulsed polarized operation is shown schematically in Fig. 5.

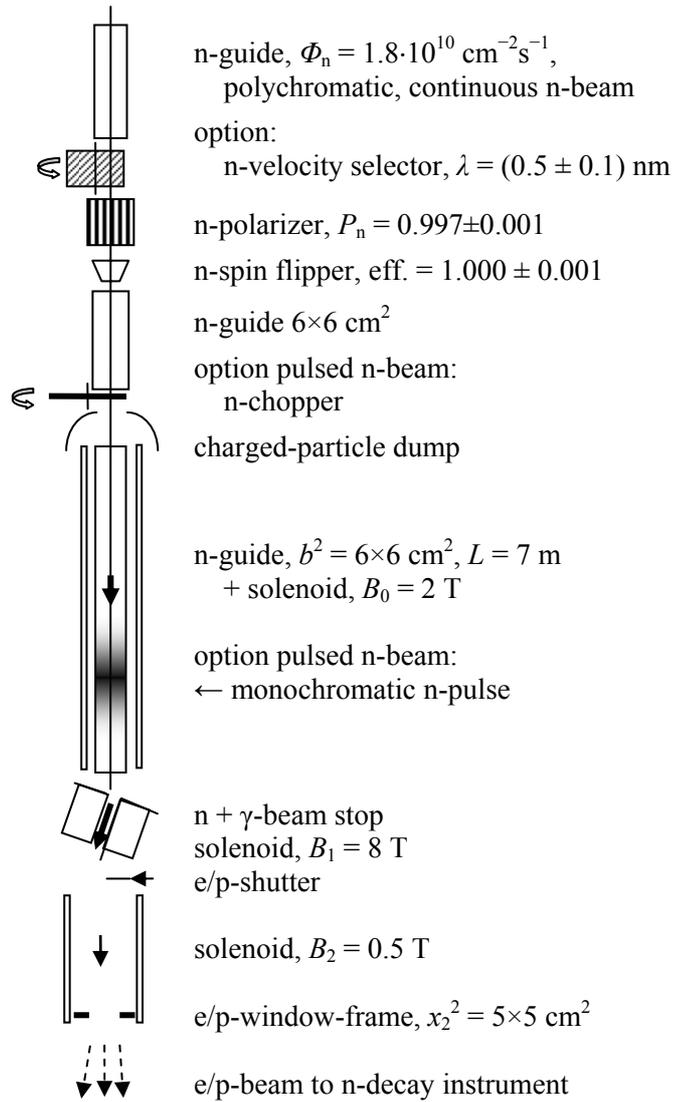

Fig. 5: The PERC beam station in 'standard configuration', equipped for pulsed polarized neutron operation. The neutron beam enters at the top of the figure, and the n-decay products emerge at the bottom of the figure. – N.B.: Most measurements with PERC can be done with an un-pulsed (continuous) n-beam.

n-guide, $\Phi_n = 1.8 \cdot 10^{10}$ cm$^{-2}$s$^{-1}$, polychromatic, continuous n-beam

option:
  n-velocity selector, $\lambda = (0.5 \pm 0.1)$ nm

n-polarizer, $P_n = 0.997 \pm 0.001$

n-spin flipper, eff. $= 1.000 \pm 0.001$

n-guide 6×6 cm$^2$

option pulsed n-beam:
  n-chopper

charged-particle dump

n-guide, $b^2 = 6 \times 6$ cm$^2$, $L = 7$ m
  + solenoid, $B_0 = 2$ T

option pulsed n-beam:
← monochromatic n-pulse

n + γ-beam stop
solenoid, $B_1 = 8$ T

e/p-shutter

solenoid, $B_2 = 0.5$ T

e/p-window-frame, $x_2^2 = 5 \times 5$ cm$^2$

e/p-beam to n-decay instrument

## 3. Typical applications of the e/p-beam

Although the further use of the PERC e/p-beam station is left to the fantasy of the user, in the following we list some possible experiments with PERC. Different types of spectrometers can be coupled to PERC:

I.   Electron energy spectroscopy,
      using an energy sensitive detector.
II.  Electron and proton momentum spectroscopy,
      using a magnetic spectrometer, e.g. like in Fig. 6.
III. Proton energy spectroscopy, and low-energy electron-spectroscopy (item 2. below)
      using an electrostatic retardation spectrometer like $a$SPECT [12].
IV.  Proton time-of-flight (TOF) spectroscopy,
      using a proton beam pulsed by an electric gate voltage.
V.   Electron helicity spectroscopy,
      using an electron helicity analyzer.

From such spectra many quantities can be derived, some of them for the first time (items 2, 3, 7, 8, and 11 below), see also [18,19] for a general discussion of measurable n-decay parameters. Most of these n-decay parameters can be derived in single-count mode with one detector:

With unpolarized neutrons:

1. e-ν correlation coefficient $a$,
   from proton spectra (energy, momentum, or TOF).
2. Fierz interference amplitude $b$,
   from electron spectra (retardation spectrometer of the $a$SPECT type).
3. Electron helicity $H_e$,
   from electron energy spectra after Mott scattering.

With polarized neutrons:

4. β-asymmetry $A$, from electron-asymmetry spectra.
5. p-asymmetry $C$, from proton-asymmetry spectra.

From the measured parameters $a$, $b$, $A$, $C$, plus from the additional information contained in the precise form of the related spectra, we obtain the following derived quantities [1-6]:

6. The anti-neutrino asymmetry $B$, from the relation $C = -0.27484\,(A + B)$ [20].
7. Weak magnetism and second class form factors $f_2$, $g_2$.
8. Radiative corrections and related QCD parameters.
9. Mass $M_R$ of right-handed $W_R$ boson.
10. Phase $\zeta$ of right-handed $W_R$ boson.
11. Scalar $g_S$ and tensor $g_T$ amplitudes, and
12. possible other 'exotic' admixtures from physics beyond the Standard Model.

Of course, even more derived quantities can be identified whenever these quantities are needed as input in other fields like astrophysics, cosmology, and particle physics [1-6].

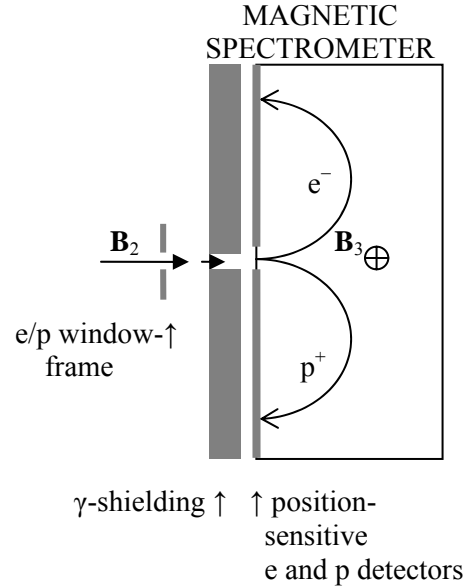

Fig. 6: Sketch of a magnetic spectrometer for neutron decay products. Electrons and protons delivered by PERC are detected in position sensitive $e^-$ and $p^+$ detectors after momentum analysis in the magnetic field.

As an example, Fig. 6 shows a possible configuration of a magnetic spectrometer coupled to PERC, which simultaneously (but not in coincidence) measures electron and proton momentum spectra via position sensitive detectors. For a spectrometer field $B_3 = 10$ mT the radius of gyration in the spectrometer will be $r_3 \leq 40$ cm. The momentum resolution of such a spectrometer scales with e/p-beam width $x_2$ and e/p-beam divergence $\theta_2$ in field $B_2$ like $\sim (x_2/r_3)\,\theta_2^2$. Reasonable values are $2x_2 = 1$ cm, and $\theta_2 = 10^0$, i.e. $B_2 = 0.22$ T for standard $B_0$, $B_1$. In this and similar applications the magnetic field $B_2$ must rapidly drop to $B_3 \ll B_2$ at the entrance of the spectrometer, such that the angular distribution of the e/p-beam will remain undisturbed. Non-adiabatic field decay will be discussed in Section 4.2.

### 4. Remaining sources of error

According to the Particle Data Group [21], at present the world-average of n-decay correlation coefficients is $a = -0.103 \pm 0.004$, $A = -0.1173 \pm 0.0013$, $B = -0.981 \pm 0.004$, and, from [10,22], $C = -0.2375 \pm 0.0024$, i.e. the relative errors are 4 %, 1 %, 0.4 % for $a$, $A$, $B$, and 1 % for $C$. With PERC we aim at a total systematic relative error below $3 \cdot 10^{-4}$ for these quantities. To have also the *statistical error* $\Delta A/A$ of the same size, the number of n-decay events must be of order $N = 1/\Delta A^2 = 10^9$, for the case of β-asymmetry $A$, and similarly for $a$ and $C$. In the standard configuration of PERC, this will require the following pure data taking times:

5 hours for continuous unpolarized,
1 day for continuous polarized ($P_n$ = 98 %),
2 days for continuous highly-polarized operation of the n-beam ($P_n$ = 99.7 %).

These times typically double when we include background and calibration measurements. For the pulsed mode of the n-beam these times will be 25 times longer (A7), but this pulsed mode will not be the typical use of PERC.

Often systematic errors are considerably larger for absolute count rate spectra $N(E)$ than they are for measured asymmetry spectra $A_{exp}(E)$, Eq. (4). For instance, the edge effect due to a thick absorbing window frame is ten times smaller for the β-asymmetry than it is for the β-count rate [22]. In most of the following, error estimates will be given for the count rate spectra and not for the asymmetry spectra. So the errors given are conservative for any prospective use of PERC. Table 1 gives a summary of PERC's error budget, which will be discussed in detail in the following sections.

Table 1: Error budget of PERC in standard configuration.
The numbering in this list is the same as in Sections 4.1 to 4.5.

| SOURCE OF ERROR | COMMENT | \| SIZE \| OF CORRECTION | SIZE OF ERROR: |
|---|---|---|---|
| (1) non-uniform n-beam | for $\Delta \Phi/\Phi$ = 10 % over 1 cm width | $2.5 \cdot 10^{-4}$ | $5 \cdot 10^{-5}$ |
| (2) other edge effects on e/p-window | for max. gyr. radius = worst case | $4 \cdot 10^{-4}$ | $1 \cdot 10^{-4}$ |
| (3) magn. mirror effect, contin's n-beam | | $1.4 \cdot 10^{-2}$ | $2 \cdot 10^{-4}$ |
|     magn. mirror effect, pulsed n-beam | for $\Delta B/B$ = 10 % over 8 m length | $5 \cdot 10^{-5}$ | $< 10^{-5}$ |
| (4) non-adiabatic e/p-transport | | $5 \cdot 10^{-5}$ | $5 \cdot 10^{-5}$ |
| (5) background from n-guide | ⎫ is separately measurable | $2 \cdot 10^{-3}$ | $1 \cdot 10^{-4}$ |
| (6) background from n-beam stop | ⎭ | $2 \cdot 10^{-4}$ | $1 \cdot 10^{-5}$ |
| (7) backscattering off e/p-window | | $1 \cdot 10^{-3}$ | $1 \cdot 10^{-4}$ |
| (8) backscattering off e/p-beam dump | | $5 \cdot 10^{-5}$ | $1 \cdot 10^{-5}$ |
| (9) backscatt. off plastic scintillator | ⎫ for method I (Sect.3) = worst case | $2 \cdot 10^{-3}$ | $4 \cdot 10^{-4}$ |
|     ~ same with active e/p-beam dump | ⎭ | – | $1 \cdot 10^{-4}$ |
| (10) neutron polarisation | present status | $3 \cdot 10^{-3}$ | $1 \cdot 10^{-3}$ |

*4.1 Errors linked to n-decay volume*

The effective *length* of the n-decay volume is a source of error only in *continuous* n-beam operation. In this case the precise length of the n-decay volume may influence the measured asymmetry signal via the e/p-magnetic mirror effect. This effect will be treated in Section 4.2. In *pulsed* n-beam operation, the length of the decay volume is unimportant as long as the n-pulse is fully contained within the uniform part of the $B_0$ field region during e/p-detection.

The effective *width* of the decay volume, on the other hand, requires careful investigation. This width is defined by the open e/p-window of width $2x_2$ and height $2y_2$, installed at field value $B_2$ (Figs. 2, 4, 5). The absorbing frame of this window is the source of errors number (1), (2), and (7) of Table 1.

(1) *Non-uniform neutron beam density*: Edge effects due to e/p-absorption on the e/p-window frame, Fig. 4, cancel for uniform n-density, but can play a role if n-density in the active volume varies across the n-beam. The e/p-radius of gyration at the position of the e/p-window is $a_2 = r_0 (B_0/B_2)^{1/2} \sin \theta_0$, with $\theta_0 \leq \theta_c$. Let the n-beam density profile in x-direction have a slope $\partial \Phi_n / \partial x_0 = \xi_\pm$ at the lateral n-beam positions $x_0 = \pm (B_2/B_0)^{1/2} x_2$. The $x_0$ are those positions within the active volume which are magnetically connected to the left and right edges of the frame. For simplicity, let the n-beam have the same profile in y-direction. Then, after integration of the effect over all e/p starting points and over all angles of emission $\theta_0 \leq \theta_c$, the cumulated edge effect due to this mismatch requires a correction of size

$$\Delta_1 = 0.3 \, (\xi_- - \xi_+) \frac{x_0^2}{x_2} \frac{B_0}{B_1} \left( \frac{B_0}{B_2} \right)^{1/2}, \qquad (6)$$

where we have used $\langle \sin^2 \theta_0 \rangle = \tfrac{1}{3}(1 - \cos \theta_c + \sin^2 \theta_c)$, which is $\approx B_0/2B_1$ for $\theta_c \ll 1$.

Even for slopes in neutron density as steep as $\xi_+ = -\xi_- = 10$ %/cm of the n-beam profiles in x- and y- direction, this correction is only $\Delta_1 = 2.5 \cdot 10^{-4}$ at the endpoint of the energy spectrum, for standard configuration of PERC. This effect can be corrected for with an accuracy of at least 20 %, so the error due to this effect will be below $5 \cdot 10^{-5}$.

(2) *Further edge effects on e/p-window*: Some of the spiraling decay particles may hit the frame of the e/p-window from 'inside', i.e. on its narrow inner face of thickness $\delta$. Concerned are only particles within a distance $2a_2$ to the window frame, i.e. a fraction $a_2/f$ (see Eq. (8) below for $f$), and of these only a fraction $\delta/\Lambda_2$ will actually hit the inner face, with the gyration length (Appendix 2)

$$\Lambda_2 = 2\pi a_2 \cos \theta_2 \approx 2\pi a_2 \text{ for } \theta_2^2 \ll 1, \qquad (7)$$

so the radius of gyration $a_2$ cancels and the correction is of size $\delta/(2\pi f)$. For frame thickness $\delta = 2$ mm, and standard configuration ($f = 2.5$ cm) this correction is about 1 % and is far too large. The remedy is to cut the inner edge of the frame obliquely such that it slightly recedes under the maximum angle of e/p incidence, which is $\theta_2 = 15°$. The inner face then can no longer be hit by the spiraling particles. However, we then still have to consider the following effect.

Some of the electrons that hit the frame of the e/p-window in a regular way, i.e. almost head-on, may scatter out through the inner face of thickness $\delta$. The mean 1/e range of β's from n-decay is 40 mg/cm² or 0.4 mm for density 1g/cm², see Fig. 2-16 in [23]. When we assume that half of the electrons that hit the frame at a distance to the edge of up to $\delta' = 0.2$ mm will leave the frame through its inner face, then their fraction is

$$\Delta_2 = \frac{\delta'}{4f}, \quad \text{with} \quad \frac{1}{f} = \frac{1}{x_2} + \frac{1}{y_2} \qquad (8)$$

of all events, i.e. $\Delta_2 = 2 \cdot 10^{-3}$ for standard window size. This fraction must be suppressed at least by a factor of 5 to about $\Delta_2 = 4 \cdot 10^{-4}$ by making the window frame 'active', using for it a plastic scintillator in anticoincidence with the e/p-detector. In our experience, 2 mm thick plastic scintillators can be built with > 90 % detection efficiency for through-going electrons, but threshold effects may require additional care. The maximum electron time-of-flight between the e/p-window and the detector will be $t_e = 0.5$ μs, hence the maximum dead time in electron counting at rate $I_s = 10^5$ s$^{-1}$ due to this veto will be $(a_2/f) I_s t_e = 4 \cdot 10^{-3}$ which is tolerable. The size of the correction $\Delta_2$ can be estimated with an accuracy of at least 25 %, so the error due to this correction will be below $1 \cdot 10^{-4}$. The effect can also be studied experimentally by varying $f$. − For protons this effect plays no role due to their small absorption length.

The above edge effects, related to the finite thickness $\delta$ of the e/p-frame, can be avoided altogether if, as already mentioned, no e/p-window but position-sensitive e/p-detection is used, with post-selection of the active e/p-beam area.

*4.2 Errors linked to e or p angle of emission*

(3) *B-field errors*: The cut-off angle $\theta_c$ of e/p-emission depends on the ratio $r = B_0/B_1$, Eq. (1), so how precisely must we know this ratio? From Eq. (A11) we find that, for standard field configuration, $|\Delta A/A| = 0.077 |\Delta r/r|$. This means that we must know the ratio $B_0/B_1$ with a precision of $10^{-3}$ to have the error in $A$ below $10^{-4}$. This is feasible if $B(z)$ is made sufficiently uniform in the region near its maximum value $B_1$.

Variations of the magnetic field $B(z)$ along the n-beam axis $z$ will change the local cut-off angle $\theta_c$ of charged particle transport to the detector, Eq. (1), and with it the measured asymmetry parameter, Eq. (4). Local depressions of $B(z)$ lead to uncontrollable trapping of decay protons and electrons and must be strictly avoided. This means that $B(z)$ must have a positive slope along $z$ all the way up to the maximum field $B_1$. If we call $<B>$ the spatial average of field $B$ over the n-decay volume (weighted with neutron density, which, however, is rather uniform within the guide), then the spatial average of the asymmetry parameter, Eq. (4), becomes

$$\frac{\langle A_{\exp} \rangle}{A_0} = \frac{1}{2} \frac{\langle B \rangle}{B_1} \frac{1}{1 - \langle \sqrt{1 - B/B_1} \rangle}, \qquad (9)$$

with $B$ below $B_1$. It turns out from this that for PERC, with a magnetic barrier $B_1 > B_0$, the sensitivity of an asymmetry parameter $A_{\exp}$ to variations of $B(z)$ is much lower than it is for PERKEO with no barrier, $B_1 = B_0$, cf. also Fig. 3c.

For *continuous* n-beam operation, the averaging in Eq. (9) must be done all along the neutron beam up to the n-beam stop, i.e. must include all the large variations in $B(z)$. We use for $B(z)$ the field of an 8 m long solenoid of 0.1 m diameter, plus Eq. (12) below for the field in the magnetic mirror region, with $B_1$ field decay length $\zeta = 25$ cm, and place the n-beam stop at the position where $B(z)$ has risen to $0.9 \cdot B_1$. We find that the magnetic mirror effect in these fields lowers the asymmetry parameter only by $\Delta_3 = <\Delta A_{\exp}/A_0> = -1.4$ % (from ½(1 + cos 30⁰) = 93.3 %, from Eq. (4), to 92.0 %, from Eq. (9)). An uncertainty in the absolute position of the n-beam stop of $\Delta z = \pm 1$ mm leads to a $\pm 1$ % (relative) uncertainty in the magnetic mirror correction $\Delta_3$. We assume that other uncertainties in $B(z)$ contribute another 1 % error, which seems feasible when the average in Eq. (9) is taken over an active n-beam volume which is well defined by the n-guide. The overall absolute error due to the magnetic mirror correction $\Delta_3$ then is $2 \cdot 10^{-4}$. Hence, magnetic mirror effects do not preclude the use of PERC in the continuous n-beam mode for absolute measurements of asymmetry parameters, and hence continuous n-beam will be the standard mode of operation of PERC.

In *pulsed* n-beam operation the detector gate will be open only as long as the neutrons are fully contained within the uniform field region of $B_0$, and the magnetic mirror correction are even smaller. For PERC we find from Eq. (9), for mean field $<B_0> = B_0(0)$ and for $B_0 << B_1$, that

$$\Delta_3 = \left| \frac{\langle \Delta A_{\exp} \rangle}{A_0} \right| \approx \frac{1}{4} \frac{(\Delta B_0)^2}{B_0 B_1}, \qquad (10)$$

with the variance of the magnetic field $\Delta B_0 = (<B_0^2> - <B_0>^2)^{½}$. For the (rather extreme) case that $B_0(z)$ rises linearly as much as 10 % over the length of the active n-beam volume, this correction is $\Delta_3 = 5 \cdot 10^{-5}$ for standard field configuration, and can be calculated with rather high accuracy. Hence, magnetic mirror effects are completely innocuous with PERC in the pulsed n-beam mode.

(4) *Non-adiabatic transport*: Non-adiabatic processes set a limit on how fast the fields $B_0$ and $B_2$ can rise or fall. Non-adiabatic processes can change the pitch angle $\theta$ of the helical e/p trajectories in an uncontrollable energy dependent way. The adiabatic condition requires that the change of the magnetic field $B(z)$ along $z$ is small during one cycle of the helical e/p motion, i.e.

$$\gamma \equiv \frac{\Lambda}{B} \left| \frac{\partial B}{\partial z} \right| << 1. \qquad (11)$$

Non-adiabatic effects are strongest in regions of low field $B(z)$ where $\theta$ is small, so $\cos \theta$ can be set to 1, and, from (7), the gyration radius is $\Lambda = 2\pi r = 2\pi p/eB$. We require that the adiabatic criterion is fulfilled equally well everywhere along the trajectory, i.e. $\gamma =$ const. Then from

$$\left|\frac{\partial B}{\partial z}\right| = \frac{e\gamma}{2\pi p} B^2 \quad \text{follows} \quad B(z) = \frac{B_1}{|z - z_0|/\zeta + 1}, \tag{12}$$

for $B(z)$ up to and beyond $B_1$, with field decay length $\zeta = 2\pi r_1/\gamma$, and $r_1 = p/eB_1$. This formula for $B(z)$ is used in Fig. 1. The length it takes for a field to decay from $B_1$ at $z_0$ to $B_2$ at $z$ then is

$$|z - z_0| = \zeta(B_1/B_2 - 1) \approx 2\pi r_2/\gamma \quad \text{for} \quad B_2 \ll B_1. \tag{13}$$

From the results of the theoretical study [24] we then find that, for $\zeta = 25$ cm and $B_1 \gg B_0$, non-adiabatic losses occur with probability

$$\Delta_4 \propto \exp(-0.2 B_0 \text{(mT)}), \tag{14}$$

i.e. $\Delta_4 = 5 \cdot 10^{-5}$ for $B_0$ and $B_2 > 50$ mT. These losses can be studied both experimentally, by varying the overall magnetic field strength, and by numerical simulation.

When a low-field device like the magnetic spectrometer in Fig. 6 with $B_3 \ll B_2$ is coupled to PERC, then the field $B_2$ is required to drop to $B_3$ in a non-adiabatic way, otherwise the transition of the charged particles to low field is not well defined. In this non-adiabatic case the field decay length must be short compared to the gyration length, i.e. $\zeta \ll 2\pi r_2$. At electron energy as low as $E = 10$ keV the gyration length is still $\zeta = 100$ mm, for $B_2 = 0.2$ mT. We have calculated the magnetic field of a rectangular solenoid of width 2 cm and height 6 cm, optimized for rapid decay of $B(z)$ at one end of the solenoid, by solving the inverse problem for the current sources of a step function for $B(z)$. We found that, with suitable correction coils, a field decay length $\zeta$ of 5 mm in the transition region from $B_2$ to $B_3$ is possible, which should be largely sufficient to guarantee non-adiabaticity. So the same or better limits can be given than for $\Delta_4$ above. The e/p-window frame of width $2x_2 = 1$ cm then must be positioned in the high field region of the solenoid where the radii of gyration are sufficiently small.

*4.3 n-beam related background*

Neutron decay data must be corrected for background. Here, too, PERC has a substantial advantage over previous spectrometers, because in PERC, contrary to earlier n-decay spectrometers, also n-beam related background can be measured separately 'off line' and hence can be corrected for precisely.

Environmental background from cosmic rays and from neighboring instruments is much less than the e/p signal rate and can be determined by simply closing the main shutter of the n-beam line (which is at typically 100 m distance upstream and therefore itself is no source of instrumental background). Environmental background therefore poses no problem as long as it is constant in time, or as long as the activity of neighboring instruments is closely monitored. Neutron-beam related background is all additional background that appears when the main n-shutter is open, i.e. sources of beam-related background are the n-beam installations from the main shutter all the way down to the n-beam stop. The isolation of beam-related background is more subtle.

We first discuss *neutron-beam related* background for the case of *continuous* n-beam operation. We must distinguish between charged and neutral background. Registered background from the walls of the n-guides will be all neutral, because secondary charged reaction products (in general electrons) released from the guide walls in field $B_0$ will gyrate back onto the guide wall and will be absorbed there. Also secondary charged background released from the back of the n-beam stop will not be registered because it will not magnetically be linked to the e/p-detector, cf. Fig. 2.

The only n-beam related *charged* background that must be considered is electrons released from the backside of the last piece of material (n-window or other) which crosses the n-beam upstream of PERC. These are prompt Compton or photo electrons from interactions with background γ-rays, or delayed β-particles after (n, γ)-activation. As PERC's magnetic field $B_0$ is coaxial with the n-beam, these electrons may be magnetically guided all the way to the e/p-detector. However, their transport to the e/p-detector will be strongly suppressed: if the residual magnetic field at the origin of the electrons is the 1 mT holding field for n-polarization, then only a fraction of $6 \cdot 10^{-5}$ will surmount the magnetic mirror field $B_1$; furthermore, in 1 mT the initial radius of gyration of such background electrons (with comparable momentum as the β's from neutron decay) is of order meters, i.e. it is much larger than the width of the n-guide, and, above some geometric critical angle of emission, these electrons will be absorbed by the nearby walls of the n-guide. Which of the two effects will be dominant depends on the geometric details of the setup. This electron background therefore should be negligible, but, if necessary, it can be investigated separately off-line by blocking the neutron beam upstream with a $^6$LiF shutter.

So we only need to consider *neutral* beam related background. In PERC we will be able to isolate all kinds of neutral background by switching off the e/p-beam, and with it the signal, without affecting neutral background. This can be done in two different ways. The first is to switch off the mirror field, $B_1 = 0$, so e/p transport to the detector is cut off while neutral background is not affected. However, as the $B_1$ magnet is superconducting, this cannot be done very often. The second method is to insert an e/p absorbing shutter just downstream of the n-beam stop, see Fig. 5, thin enough so that the neutral background coming through the n-beam stop is not significantly attenuated. However, again, the neutral background may release secondary electrons from the back of this e/p-shutter, so background would be over-corrected. This small secondary background, however, can separately be measured by setting the e/p-transport field $B_2$ to zero. We conservatively assume that separate measurement of n-beam related neutral background is possible with 5 % accuracy.

Although all n-beam related background in PERC can separately be measured, we still want to estimate its magnitude beforehand. With PERC, n-beam related background will be strongest for case I in the list of Section 3, i.e. an energy sensitive detector with direct view onto the n-beam stop. Therefore the following discussion will be limited to this latter 'worst case'. Our experience with PERKEO shows that background from the n-polarizer is considerably smaller than background from the internal n-guide and from the n-beam stop, and we therefore will discuss only these two background sources.

(5) *Background from the neutron guide system*: In PERC, background due to n-reflection losses on the walls of the guide plays a dominant role because these losses take place also within the active decay volume. In the average, a 'supermirror' n-guide has $\delta_n = 3$ % neutron losses per bounce [25]. For a square cold guide of width $b = h = 6$ cm, the measured divergence $\theta_n$ of the (unpolarized) n-beam [26] leads to a mean free path between successive wall bounces of $\Lambda_n \approx 8/(\pi\theta_n)(b^{-1}+h^{-1}) = 12$ m. For a given neutron intensity $I_n$, the number of bounces over a length d$z$ of the guide is $I_n \mathrm{d}z/\Lambda_n$. About 80 % of the lost neutrons are captured by boron nuclei in the glass substrate of the guide, under emission of a prompt 0.5 MeV γ-ray, while about 20 % of the lost neutrons are captured in the Ni and Ti layers of the supermirror coating of the guide, under emission of 6 to 9 MeV γ's.

In PERC's standard configuration, the e/p-detector is protected from the active n-decay volume by the n-beam stop, see Fig. 2. The γ-transmission of a 5 cm thick layer of lead in the n-beam stop is $T_\gamma = 1\cdot 10^{-3}$ at 0.5 MeV, and $T_\gamma = 0.1$ at 6 to 9 MeV γ-energy. More lead probably will not lower these numbers because of the oblique open channel in the n-beam stop through which the charged decay particles are guided. If an organic phosphor of thickness 5 mm is used as e/p-detector, its γ-sensitivity is $\varepsilon_\gamma = 5$ % at 0.5 MeV and 1 % at 6 to 9 MeV, for a discriminator threshold below 50 keV [27].

A detector installed at field value $B_2$ must have an area of size $\sim b^2 B_0/B_2$. The *background γ-intensity* $I_\gamma$ in this detector, coming from the guide, regarded as a line source, is integrated to

$$\frac{I_\gamma}{I_n} = \frac{c}{z_0} - \frac{c}{z_0 + L}, \quad \text{with} \quad c = \frac{b^2 \delta_n \varepsilon_\gamma T_\gamma}{4\pi \Lambda_n} \frac{B_0}{B_2}. \tag{15}$$

Here $L$ is the length of the decay volume and $z_0$ the distance between n-beam stop and e/p-detector. With the relative *signal intensity* $I_s/I_n$ of the e/p-beam taken from (A3) and (A2), the background from the n-guide becomes

$$\varDelta_5 = \frac{I_\gamma}{I_s} = \frac{I_\gamma}{I_n} \frac{I_n}{I_s}. \tag{16}$$

For $z_0 = 2$m and standard field configuration, $I_\gamma/I_n = 3\cdot 10^{-10}$ for the sum of γ-transitions of an infinitely long n-guide ($L \to \infty$). Hence, with $I_s/I_n = 1.5\cdot 10^{-7}$, the background ratio is $\varDelta_5 = 2\cdot 10^{-3}$. With $I_\gamma$ separately measurable with 5 % accuracy, $\varDelta_5$ contributes an error of $1\cdot 10^{-4}$. $\varDelta_5$ is roughly independent of the n-beam mode employed (continuous/pulsed or polarized/unpolarized). When a supermirror polarizer is used, the divergence of the n-beam usually is somewhat larger than with an unpolarized n-beam, and $\Lambda_n$ accordingly somewhat shorter.

(6) *Background from the n-beam stop:* Background from the $^6$LiF n-beam stop are γ-rays from the (n, γ) and fast neutrons from the (α, n) reaction after $^6$Li (n,α) neutron capture, each with probability of $1\cdot10^{-4}$ [28], so the total background from the n-beam stop is

$$\Delta_6 = (I_\gamma' + I_{\text{fast n}})/I_s. \tag{17}$$

Let the plastic scintillator be positioned at 2 m distance from the n-beam stop (solid angle $\Omega_\gamma \sim 2.5\cdot10^{-3}$). Its detection efficiency is $\varepsilon_\gamma \sim 2\%$ and 1 % for the 1.6 MeV and 5 to 7 MeV fluorine γ-transitions, and $T_\gamma \sim 10^{-1}$. Then $I_\gamma'/I_n = 1\cdot10^{-4}\cdot\varepsilon_\gamma T_\gamma \Omega_\gamma/4\pi \sim 2.5\cdot10^{-11}$, and $I_\gamma'/I_s \sim 1.7\cdot10^{-4}$. The fast neutrons have energies 3 to 10 MeV. A layer of 10 cm of water equivalent added to the beam stop lowers their number by 3 [29]. The detector efficiency for fast neutrons being 2 % in this energy range [30], we obtain a background to signal rate of $I_{\text{fast n}}/I_s \sim 3\cdot10^{-5}$. The error due to the correction $\Delta_6 = 2\cdot10^{-4}$ then is $1\cdot10^{-5}$.

A fraction of cold neutrons is backscattered from the $^6$LiF n-beam stop. In $^6$LiF the s-wave scattering cross section $\sigma_s$ is small compared to the absorption cross section $\sigma_a$, namely $\sigma_s/\sigma_a = 5\cdot10^{-3}$. In first order of $\sigma_s/\sigma_a$, this backscattered fraction is $0.15\cdot\sigma_s/\sigma_a = 8\cdot10^{-4}$. For the integration over depth $x$ of the beam stop, we have used

$$\frac{1}{2}\int_0^\infty e^{-x}\int_0^1 e^{-x/t}\,\mathrm{d}x\mathrm{d}t = 0.15, \tag{19}$$

with $t = \cos\theta$ and n-scattering angle $\theta$. These backscattered neutrons are absorbed in the boron of the guide, and their contribution to the background from the primary neutrons, given by $\Delta_5$ above, can be neglected.

In *pulsed* n-beam operation, background from the n-beam stop plays no role because the gate of the e/p-detector will be closed before the n-pulse arrives at the beam stop. The periodic n-chopper, on the other hand, is an additional source of background, because the n-chopper closes the n-beam while the gate of the e/p-detector is open. This background can be measured separately with the n-beam open and the chopper permanently stopped in 'closed' position. Alternatively, it can be determined together with all other beam related background in the ways described above for the continuous n-beam case. Anyway, as the distance of the e/p-detector to the n-chopper is about five times greater than its distance to the n-beam stop, the background from the n-chopper is 25 times lower than the background $\Delta_6$ from the n-beam stop in the continuous n-beam mode derived above, and hence will be negligible.

Finally, in pulsed n-mode, the background rate $\Delta_5$ from the neutron guide will depend on the momentary position of the moving neutron cloud, i.e. will be time dependent, while the signal rate $I_s$ from the same cloud will remain constant in time. This will give us another handle to identify such residual beam related background.

*4.4 e/p-backscattering*

There are several objects on which the decay electrons and protons can backscatter. The e/p will all impinge near to normal onto the surfaces of these objects, because these are all positioned in low magnetic field. Backscattering effects can be corrected for with a tested precision of 10 % [31,32].

(7) *Backscattering off the e/p-window* go back to the e/p-beam dump and, at first sight, are no problem. They can, however, disturb the measurement when they are magnetically reflected by the $B_1$ mirror field and return to the frame, but miss the frame on their second passage. For $B_2 \ll B_1$ the joint probability for this event is calculated to

$$\Delta_7 = \frac{8\eta}{3\pi^2}\frac{r_0}{f}\frac{B_0}{\sqrt{B_1 B_2}}. \tag{20}$$

The electron backscattering coefficient $\eta$ for normal incidence on a plastic scintillator is 5 % for electron energies below 100 keV [32] and diminishes to half this value at the β-endpoint energy, see [33] and Table 10.1 in [23], while $r_0$ increases with energy. For our standard field configuration the correction then is below $\Delta_7 = 7 \cdot 10^{-5}$ everywhere in the spectrum. In Section 4.1 an active frame was required for the suppression of $\Delta_2$, which will suppress $\Delta_7$ further to at least $2 \cdot 10^{-5}$. When calculated with 50 % accuracy the error from $\Delta_7$ will be $1 \cdot 10^{-5}$. For proton backscattering, the effect is even smaller [34].

(8) *Backscattering off the e/p-dump*: some effective dump must be foreseen to eliminate the large fraction of decay particles emitted or reflected into 'upstream' direction, i.e. back towards the reactor source. From (A10), a fraction $B_0/4B_1 = 1/16$ of all decay particles reaches the detector, while the remainder $1 - B_0/4B_1 = 15/16$ goes to the dump. To effectively dump these particles the n-guide is interrupted over a gap-length of ~0.3 m at the entrance of the $B_0$ solenoid. At this distance an e/p-beam dump of low $Z$ with back-scattering coefficient $\eta$ is installed at right angles to the field lines, as indicated in Fig. 5. If backscattering is isotropic then only a fraction $B_s/2B_1$ of the backscattered electrons will surmount the magnetic field barrier $B_1$ and reach the e/p-detector. Altogether the particles backscattered from the dump reaching the detector contribute to the e/p count rate a fraction

$$\Delta_8 = \frac{\eta}{2} \frac{B_s}{B_1} \left( 4 \frac{B_1}{B_0} - 1 \right). \tag{21}$$

The stray field of the $B_0$-solenoid drops to the value of $B_s \sim 1$ mT over a distance of several coil diameters, i.e. $B_s/2B_1 \sim 6 \cdot 10^{-5}$. With $\eta = 5$ % we obtain $\Delta_8 = 5 \cdot 10^{-5}$ for standard field configuration. This correction can be calculated with 20 % precision, and the error is $1 \cdot 10^{-5}$. Proton backscattering can in principle also be prevented by a −1 kV electric potential. The number of particles backscattered from the primary n-guide on the upstream side of the gap, or from the chopper, Fig. 5, is negligible.

(9) *Backscattering off the e/p-detector*: In proton time-of-flight or e/p magnetic spectrometry (Fig. 6), backscattering off the detector plays a minor role, as long as the particle is detected at all, because the e/p must only be registered but not analyzed in the detector. Backscattering off the e/p-detector therefore is relevant only for spectroscopic method I (Section 3), i.e. with an energy sensitive detector installed in field $B_2$. Of all e/p backscattered off this detector, only those emitted into a small solid angle $\Delta\Omega/2\pi \approx B_2/2B_1 = 3 \cdot 10^{-2}$ will pass the mirror field $B_1$ and are really lost, the others are reflected back onto the detector and deposit their full energy. The backscattering rate then is

$$\Delta_9 = \pi \frac{\partial \eta}{\partial \Omega} \frac{B_2}{B_1}. \tag{22}$$

Proton backscattering can be prevented by having the detector on a −1 kV potential. Low energy electron backscattering with near to normal incidence (in low $B$-field) and near to normal back emission (magnetic mirror barrier $B_1$) occurs with $\partial\eta/\partial\Omega = 1.5$ % sr$^{-1}$ for a plastic scintillator [32] and 6 % sr$^{-1}$ for silicon [31]. Averaged over all n-decay electron energies, we expect $\partial\eta/\partial\Omega = 1$ % sr$^{-1}$ and 4 % sr$^{-1}$, respectively. For electron backscattering then $\Delta_9 = 2 \cdot 10^{-3}$ for a plastic scintillator, and $8 \cdot 10^{-3}$ for a silicon detector, for standard field configuration. Electron backscattering off the detector can be studied experimentally by varying $B_1$. It can be corrected by numerical simulation with 20 % accuracy, which gives an error of $4 \cdot 10^{-4}$ for electron backscattering off a plastic scintillator. The fraction of backscattering events that go undetected, due to a finite detector threshold, was studied experimentally with PERKEO's plastic scintillators in [35] and was found negligibly small.

One may want to further suppress electron backscattering off the detector by making the e/p-dump, which intercepts 98 % of all backscattered electrons, 'active', i.e. from a plastic scintillator. The only role of the active e/p-beam dump would be to recover the lost energy of backscattered electrons in coincidence with the primary electron detector, so it can be operated at a relatively high background (= $bgd$). That this recovery is feasible was proven in the PERKEO instrument. For a maximum electron time-of-flight across PERC of $t_e = 2$ μs, the maximum usable count rate of PERC will be severely limited by the dead time condition $(n_\beta + bgd) 2t_e \ll 1$, cf. Eq. (A2), or $n_\beta + bgd \ll 5 \cdot 10^5$, which excludes the use of PERC at the highest possible count rates.

Protons can be detected in the same plastic scintillator as the electrons by using the secondary electrons released by protons hitting a carbon foil at high negative potential [9,10]. In this case backscattering is not relevant because the electrons are not used for energy analysis. When protons are detected in a silicon detector they may suffer losses in the gold window of the detector which may be difficult to quantify, so an energy sensitive silicon p-detector used with PERC will face the same problems as it does when used with existing spectrometers.

*4.5. Neutron polarization*

(10) *Non-perfect n-polarization* $P_n$ at present requires a polarization correction
$$\Delta_{10} = 1 - |P_n| = 3 \cdot 10^{-3}, \qquad (23)$$
with an error of $1\cdot10^{-3}$ [17]. When PERC is used with polarized neutrons, neutron depolarisation during reflection from the guide walls within the active n-decay volume may further diminish $P_n$. This depolarization may take place in the magnetic stray fields from the ferromagnetic nickel layers in the Ni/Ti supermirrors coating. Therefore, for PERC's $6\times6$ cm$^2$ internal n-guide (Fig. 5), nonmagnetic supermirror coatings must be used. Such coatings have been developed in the form of NiMo/Ti multi-layers [36,37]. Supermirrors based on nonmagnetic materials like Cu instead of Ni can also be considered. We have performed test measurements on NiMo/Ti mirrors [15] which show that indeed we can exclude neutron depolarization on the level of $1\cdot10^{-5}$. This limit holds for the diagonal element of the polarization transfer matrix. The depolarization due to the off-diagonal element was also found to be zero, but only on the level of $2\cdot10^{-2}$, due to a lack of n-beam-time. This measurement will be improved later on. For the time being we assume the $1\cdot10^{-3}$ error from the polarization correction (23) to be the limit in absolute polarized neutron work with PERC.

## 5. Conclusion

With PERC a new avenue is opening up in neutron decay spectroscopy. The PERC facility will deliver not neutrons but a well defined and extremely strong beam of electrons and protons from neutron decay. This beam can feed a variety of dedicated instruments, to be installed by various successive users, for many different experimental investigations, as listed in Section 3. The huge gain in phase space density, as compared to previous neutron decay spectrometers, is partly used to suppress statistical errors, and partly to suppress systematic errors, as is discussed in Sections 2.4 and 4. All experimental errors, like beam-related background, e/p-backscattering, or e/p-magnetic mirror effects, listed in Table 1, are shown to be limited to $2\cdot10^{-4}$ or better, ten times lower than in presently running spectrometers. The only exception, in the case of *polarized* neutron decay, is the present error in neutron polarization of $<1\cdot10^{-3}$, which, however, is continuously being improved, independent of the PERC project.

**Appendix 1: Expected count rates with PERC**

The cold-neutron beam station PF1B at the 'ballistic' supermirror guide H113 [25,26] of ILL Grenoble has an average neutron capture flux (thermal equivalent flux) $\Phi_n = 1.8 \cdot 10^{10}$ cm$^{-2}$s$^{-1}$ (as measured in summer 2006 after an upgrade of H113) over a rectangular cross section $bh = 6 \times 20$ cm$^2$. The total neutron intensity is $I_{n0} = bh\Phi_n = 2.2 \cdot 10^{12}$ s$^{-1}$. Similar fluxes are now available at FRM II in Munich.

The in-beam neutron β-decay rate is $n_{\beta 0} = (L/v_0\tau)I_{n0}$, with neutron lifetime $\tau = 885$ s, $v_0 = 2200$ ms$^{-1}$, and length $L$ of the decay volume. The decay rate per length of guide is $n_{\beta 0}/L = 1.1 \cdot 10^6$ s$^{-1}$m$^{-1}$. In the proposed installation, the decay volume will have a cross-section $b^2 = 6 \times 6$ cm$^2$, and with a 20 % loss due to the gaps in the guide, the neutron intensity becomes

$$I_n = 0.8 b^2 \Phi_n = 5 \cdot 10^{11} \text{s}^{-1}. \tag{A1}$$

For an active volume of length $L = 7$ m, the in-beam n-decay rate from a *continuous unpolarized* n-beam then is

$$n_\beta = \frac{L}{v_0 \tau} I_n = 3.6 \cdot 10^{-6} I_n = 1.9 \cdot 10^6 \text{s}^{-1}. \tag{A2}$$

One half of the decay particles are emitted 'downstream' towards the end of the guide, and of these a fraction $1 - (1 - B_0/B_1)^{\frac{1}{2}}$, which is $\approx B_0/2B_1$ for $B_1 \gg B_0$, passes the magnetic field barrier $B_1$ (Appendix 2). The signal intensity $I_s$ of the decay electrons (protons), which form a beam of initial cross section given by the e/p-window as $4x_0 y_0 = 5 \times 5$cm$^2$, then is given by

$$I_s = \frac{1}{2} \frac{4 x_0 y_0}{b^2} \frac{B_0}{2 B_1} n_\beta = 0.043 n_\beta = 8 \cdot 10^4 \text{s}^{-1}, \tag{A3}$$

for standard field configuration $B_0/B_1 = \frac{1}{4}$, and a detector efficiency near 100 %. In the continuous mode the n-beam will be used preferentially with neutrons polarized to 98 %, with polarizer transmission $T_n = \frac{1}{2} \cdot 0.4 = 0.2$ (with respect to the *unpolarized* incoming n-beam). The count rate then is, for *continuous polarized* n-beam operation, $T_n \cdot I_s = 1.6 \cdot 10^4$ s$^{-1}$.

For *pulsed* neutron beam operation, the neutron wavelength spectrum must first be limited to a band $\Delta\lambda = \lambda_2 - \lambda_1$ of intensity $I_n'$. With n-velocity $v$(ms$^{-1}$) $= 395/\lambda$(nm) for the mean wavelength $\lambda = \frac{1}{2}(\lambda_1 + \lambda_2)$, and initial neutron pulse length $\Delta L = v \Delta t$, the intensity $I_s'$ of decay particles from the pulsed n-beam then is derived from the intensity $I_s$ from the continuous n-beam as

$$I_s' = \frac{I_n'}{I_n} \frac{v_0}{v} \frac{\Delta L}{L} \frac{\Delta t'}{T} I_s, \tag{A4}$$

with chopper opening time $\Delta t$, detector gate time $\Delta t'$, and chopper period $T$.

The decay volume of length $L' < L$ for the pulsed beam must be entirely within the uniform region of field $B_0$. Let $\Delta z_1$ be the distance from the chopper to the start of the decay volume, and $\Delta z_2$ the distance from the end of the decay volume to the n-beam stop, cf. Fig. 5. To be specific, we set $\Delta z_1 = \Delta z_2 = 0.6$ m, $L' = 6$ m, and the upper wavelength $\lambda_2 = 0.6$ nm, i.e. the lower velocity $v_2 = 660$ ms$^{-1}$. The chopper period then is

$$T = \frac{L' + \Delta z_2}{v_2} = 10 \text{ms}. \tag{A5}$$

When we make the realistic assumption that the pulsed neutron intensity $I_n'$ is linear in bandwidth $\Delta\lambda$, then from elementary kinematics the count rate $I_s'$ of decay particles is found to be at maximum when chopper and detector opening times $\Delta t$, $\Delta t'$, and neutron band width $\Delta\lambda$ are:

$$\Delta t = \Delta t' = \frac{L'}{3 v_2} = 3 \text{ms}, \quad \Delta\lambda = \frac{1}{3} \frac{L'}{L' + \Delta z_1} \lambda_2 = 0.2 \text{nm}, \tag{A6}$$

i.e. $\lambda_1 = \lambda_2 - \Delta\lambda = 0.4$ nm, $v = 780$ ms$^{-1}$, and $\Delta L = 2.3$ m. For short "dead volumes" $\Delta z_1, \Delta z_2 \ll L'$, this simplifies to $\Delta t = \Delta t' \approx \frac{1}{3}T$, $\Delta\lambda \approx \frac{1}{3}\lambda_2$, and $\Delta L \approx \frac{1}{3}L'$. A typical cold n-guide has about 30 % n-flux in the wavelength band $\Delta\lambda$ from 0.4 nm to 0.6 nm. For a typical velocity selector with triangular transmission function we should take half this value, i.e. $I_n'/I_n \approx 0.15$. Hence the pulsed count rate $I_s'$ is derived from the continuous count rate $I_s$ as

$$I_\text{s}' = \frac{1}{9} \frac{I_\text{n}'}{I_\text{n}} \frac{v_0}{v_2} \frac{L'}{L} \frac{L'}{L'+\Delta z_2} I_\text{s} = 0.043 I_\text{s} = 3.5 \cdot 10^3 \, \text{s}^{-1}. \tag{A7}$$

When we further account for a 20 % loss in the neutron velocity selector and a 20 % loss due to finite chopper opening/closing times, then the electron (proton) count rates from the *pulsed unpolarized* n-beam become $I_\text{s}' = 0.028 \, I_\text{s} = 2.2 \cdot 10^3 \, \text{s}^{-1}$. A *pulsed polarized* n-beam will primarily be used for absolute correlation measurements at the highest achievable neutron polarization of $(99.7 \pm 0.1)$ % with transmission $T_\text{n}' = \tfrac{1}{2} \cdot 0.2 = 0.1$ [17], so $T_\text{n}' I_\text{s}' = 220 \, \text{s}^{-1}$.

**Appendix 2: Charged particles in magnetic fields**

For neutron decay, the electron endpoint energy is $E_{\beta \, \text{max}} = 782$ keV, the proton endpoint energy is 0.75 keV, and momentum $p = c^{-1}(E^2 + mc^2 \cdot E)^{1/2}$ is limited to $p_\text{max} = 1.20$ MeV/$c$ for both protons and electrons. Let the electron/proton be emitted in field $B_0$ in downstream direction, i.e. under an angle $\theta_0 < 90^0$. The particle moves on a helix with pitch angle $\theta_0$ and radius of gyration $a_0 = r_0 \sin \theta_0$, where $r_0 = p/eB_0$ is the radius of gyration under emission at right angles to $B_0$. For $B = 1$ T, the maximum radius of gyration at $E_{\beta \, \text{max}}$ is $r_{0 \, \text{max}} = 4$ mm. While the particle makes one full turn it travels a distance $\Lambda = 2\pi a_0 \cos \theta_0$, called the gyration length.

When the particle moves against a mirror field $B_1 > B_0$, it will reach $B_1$ under an angle $\theta_1 > \theta_0$, with $\sin \theta_1 = (B_1/B_0)^{1/2} \sin \theta_0$ as long as $\theta_0 < \theta_\text{c}$, otherwise it will be reflected by the magnetic mirror effect before reaching $B_1$. The cut-off angle $\theta_\text{c}$ for reflection is

$$\sin \theta_\text{c} = \sqrt{B_0 / B_1} \tag{A8}$$

which ensures $\sin \theta_1 \leq 1$. Hence, the particles emitted under $\theta_0 < \theta_\text{c}$ which can surmount the mirror field $B_1$ will have their initial radius of gyration in $B_0$ limited to $a_0 = r_0 \sin \theta_0 \leq r_0 (B_0/B_1)^{1/2}$. At field $B_1$, the gyration radius is limited to $a_1 = a_0 (B_0/B_1)^{1/2} \leq r_0 B_0/B_1 = p/eB_1 = r_1$.

The particles that have passed $B_1$ will reach the field $B_2 < B_1$ under angle $\theta_2$, with $\sin \theta_2 = (B_2/B_0)^{1/2} \sin \theta_0$, and with radius $a_2 = a_0 (B_0/B_2)^{1/2}$. These values are limited to

$$\sin \theta_2 \leq \sqrt{B_2 / B_1}, \quad a_2 \leq r_0 B_0 / B_2 \tag{A9}$$

Of all particles emitted isotropically into one half-space (defined by $\boldsymbol{p} \cdot \boldsymbol{B}_0 > 0$), a fraction

$$N / N_0 = 1 - \cos \theta_\text{c} \approx B_0 / 2B_1 \quad \text{for} \quad \theta_\text{c}^2 \ll 1 \tag{A10}$$

will pass the magnetic mirror field $B_1 > B_0$. For asymmetric particle emission of type $W(\theta) = 1 + A_0 \cos \theta$, the measured asymmetry parameter depends on $\theta_\text{c}$ as

$$\begin{aligned} A_\text{exp} &= A_0 (1 + \cos \theta_\text{c}) / 2 = A_0 (1 + \sqrt{1 - B_0/B_1}) / 2 \\ &\approx A_0 (1 - B_0 / 4B_1) \quad \text{for} \quad \theta_\text{c}^2 \ll 1. \end{aligned} \tag{A11}$$

The figure of merit of a measurement is proportional to the inverse beam-time to reach a certain level of relative error. For an asymmetry measurement at count rate $N$ the inverse beam-time scales as

$$N A_\text{exp}^2 \propto \left(1 + \sqrt{1 - B_0 / B_1}\right)(B_0 / B_1), \tag{A12}$$

which we take as figure of merit, as displayed in Fig. 3d.